\documentclass[1p,sort&compress]{elsarticle}
\makeatletter
\def\ps@pprintTitle{%
 \let\@oddhead\@empty
 \let\@evenhead\@empty
 \def\@oddfoot{\rightline{\thepage}}
 \let\@evenfoot\@oddfoot}
\makeatother

\usepackage{graphicx}
\usepackage[colorlinks,breaklinks]{hyperref}
\usepackage{paralist}
\usepackage{listings}

\begin{document}
%
% paper title
% can use linebreaks \\ within to get better formatting as desired
\title{The RECIPE Approach to Challenges in Deeply Heterogeneous High Performance Systems}
\let\thefootnote\relax\footnotetext{DOI:10.1016/j.micpro.2020.103185 \\© 2020 Elsevier. This manuscript version is made available under the CC-BY-NC-ND 4.0 license http://creativecommons.org/licenses/by-nc-nd/4.0/ }

% author names and affiliations
% use a multiple column layout for up to two different
% affiliations

\author[1]{Giovanni Agosta}
\ead{agosta@acm.org}

\author[1]{William Fornaciari}
\ead{william.fornaciari@polimi.it}

\address[1]{\textit{Politecnico di Milano} Milano, Italy}

\author[2]{David Atienza}
\ead{david.atienza@epfl.ch}
\address[2]{\textit{\'{E}cole Polytechnique F\'{e}d\'{e}rale de Lausanne} Lausanne, Switzerland }

\author[3]{Ramon Canal}
\ead{ramon.canal@bsc.es}
\address[3]{\textit{Barcelona Supercomputing Center} and \textit{Universitat Polit\`{e}cnica de Catalunya} Barcelona, Spain}

\author[4]{Alessandro Cilardo}
\ead{acilardo@unina.it}
\address[4]{\textit{University of Naples Federico II / CeRICT} Napoli, Italy}

\author[5]{Jos\'{e} Flich Cardo}
\ead{jflich@disca.upv.es}
\address[5]{\textit{Universitat Polit\`{e}cnica de Val\`{e}ncia} Val\`{e}ncia, Spain}

\author[5]{Carles Hernandez Luz}
\ead{carherlu@upv.es}

\author[6]{Michal Kulczewski}
\ead{kulka@man.poznan.pl}
\address[6]{\textit{Pozna\'{n} Supercomputing and Networking Center} Pozna\'{n}, Poland}

\author[1]{Giuseppe Massari}
\ead{giuseppe.massari@polimi.it}

\author[5]{Rafael Tornero Gavil\'{a}}
\ead{ratorga@disca.upv.es}

\author[2]{Marina Zapater}
\ead{marina.zapater@epfl.ch}

\begin{abstract}
RECIPE (REliable power and time-ConstraInts-aware Predictive management of heterogeneous Exascale systems)
is a recently started project funded within the H2020 FETHPC programme,
which is expressly targeted at exploring new High-Performance Computing (HPC) technologies. 
RECIPE aims at introducing a hierarchical runtime resource management infrastructure
to optimize energy efficiency and minimize the occurrence of thermal hotspots,
while enforcing the time constraints imposed by the applications
and ensuring reliability for both time-critical and throughput-oriented computation
that run on deeply heterogeneous accelerator-based systems.
This paper presents a detailed overview of RECIPE,
identifying the fundamental challenges as well as the key innovations addressed by the project. %,
%which span run-time management, heterogeneous computing architectures,
%HPC memory/interconnection infrastructures, thermal modelling, reliability, programming models, and timing analysis.
%For each of these areas, the paper describes the relevant state of the art as well as
%the specific actions that the project will take to effectively address the identified technological challenges.
In particular, the need for predictive reliability approaches to maximize hardware lifetime and guarantee application performance is identified as the key concern for RECIPE, and is addressed via hierarchical resource management of the heterogeneous architectural components of the system, driven by estimates of the application latency and hardware reliability obtained respectively through timing analysis and modelling thermal properties, mean-time-to-failure of subsystems.
We show the impact of prediction accuracy on the overheads imposed by the checkpointing policy, as well as a possible application to a weather forecasting use case.
\end{abstract}


% make the title area
\maketitle

\section{Introduction and long-term objectives}
\label{intro-sec}
The current increase in available data from a variety of sources and the emergence of new application domains that require massive amount of computational power is transforming the landscape of High-Performance Computing (HPC).
HPC has always been an important infrastructure for several industrial fields (e.g., oil \& gas, finance) as well as for weather forecasting and for the advancement of physics and chemistry. 
Nowadays, it can be considered a vital infrastructure for both nations and industrial players, as testifies by the massive investments that are leading towards the construction of Exascale systems.

Emerging application domains such as computational biochemistry and high-performance machine learning, coupled with the increasing heterogeneity of the underlying hardware architectures, require new technologies able to deal with increasingly complex systems, while preserving the performance levels that HPC users can currently leverage.

In sight of the Quality of Service (QoS) needs of these emerging application domains, as well as the need to manage computational resources in a more flexible way to address the partitioning of heterogeneous resources among applications exhibiting widely different ability to exploit specific accelerators, tightly coupled approaches
for application development and management of HPC resources and energy are deemed necessary~\cite{sra2017}.
Other key challenges include the management of thermal issues through a balanced approach involving both hardware and software support, and the capability to predict failure occurrences.

The ``REliable power and time-ConstraInts-aware Predictive management of heterogenous Exascale systems'' (RECIPE) project targets such emerging HPC scenarios, characterized by new types of requirements, including time criticality and reliability,
as well as new delivery models, e.g. cloud-based solutions~\cite{fortissimo}, potentially open to a large audience of HPC users. 
These emerging scenarios require that new HPC platforms are able to support multiple applications running concurrently, possibly with conflicting Quality of Service
requirements~\cite{agosta2018managing,flich2016mango,flich2018mango,samos2019, pupykina2017optimizing}.
In addition, at the technology level, RECIPE addresses deep heterogeneity, based on dedicated accelerators like GPUs and FPGAs,
as an enabling factor for improved energy efficiency, building on the results collected from previous research
projects~\cite{flich2015mango, Zanella2018back}.

The main focus in RECIPE is on resource management \cite{Fornaciari2018reliable},
targeted at performance predictability and reliability for HPC applications.
To address current limitations, RECIPE will take a hierarchical approach to develop a Run-Time Management System (RTMS)
partitioned in two layers, a {\em global} resource manager providing proactive fault tolerance and coarse-grained resource allocation,
as well as a {\em local} resource manager, ensuring reactive fault tolerance and fine-grained resource allocation.
The fine-grained resource management and predictive strategies will support the combined optimization of reliability and timing.
The interface between the runtime management system and the application will be mediated by the programming model,
supporting the expression of application expectations.
Furthermore, the runtime management system will rely on hardware monitors and a hardware abstraction layer,
providing a uniform view of the available accelerators, including GPU, manycore, and FPGA devices,
and exposing partitioning mechanisms used to ensure scalability, system-wide reconfigurability, and dedicated resource provisioning.

The RECIPE project aims at providing two key contributions.

\paragraph{A global manager and proactive strategies for reliability, energy efficiency and QoS} 
The \emph{global resource manager} operates at the \emph{system level}, and aims at increasing the system efficiency in terms of performance/watt, due to power / performance / thermal aware allocation; reducing the mean time to failure (MTTF) due to proactive strategies; and enforcing timing guarantees for critical applications.

\paragraph{A local manager and reactive strategies for reliability, energy efficiency and QoS}
The \emph{local manager} operates at the level of the \emph{individual heterogeneous node} (typically composed of one or more CPUs as well as heterogeneous accelerators) aims at improving the overall utilization of non-CPU computing devices under heavy condition of massively parallel workload;
improving the Energy-Delay product with respect to current state-of-the-art HPC infrastructures;
reducing the probability of faulty executions;
reducing the average recovery time with respect to classical checkpoint-restart approaches;
and enforcing timing guarantees for critical applications in cooperation with the \emph{global resource manager}.

To achieve the aforementioned goals RECIPE is developing technologies focusing on:

\begin{description}
\item[Fault prediction, prevention and recovery] 
Through predictive techniques, we aim at improving the error detection coverage to provide more precise estimates of MTTF to the resource management infrastructure. 

\item[Thermal modelling technique]
By developing thermal models of the system, we aim at reducing thermal gradients, thus improving system MTTF due to thermal stress. 

\item[Timing analysis and enforcement]
Through more precise timing analysis and enhanced QoS enforcing at the run-time manager, we aim at achieving zero deadline misses for critical applications under the targeted failure rates. 

\item[Runtime manager integration with hybrid programming models]
By coupling the resource manager with hybrid programming models, we aim at radically reducing the integration cost, as well as simplifying the programming in such hybrid models. 

\item[Hardware abstraction layer for resource disaggregation]
Through an appropriate disaggregation layer allowing remote access to heterogeneous accelerators, we aim at increasing the effective usage of available resources under non-saturated conditions as well as identifying optimal matches of application requests and available resources.

\end{description}

\paragraph{Organization of the paper}
The rest of this paper is organized as follows.
In section~\ref{background-sec} we review the technological background, while in section~\ref{architectural-sec} we provide an overview of the deeply heterogeneous architecture upon which the RECIPE technologies are built, as well as the hardware abstraction layer that provides resource disaggregation. 
In section~\ref{rtms-sec} we describe the key innovations provided by the RECIPE project in terms of predictive resource management and its integration with the programming model, while in section~\ref{usecases-sec} we discuss the application scenario, with an in-depth analysis of one of them.
In section~\ref{conclusions-sec} we draw some conclusions and highlight future research directions.

\section{Background and open issues}
\label{background-sec}

This section describes the current technological background which is directly relevant for RECIPE.
This spans several areas including run-time management, heterogeneous computing architectures,
HPC memory/interconnection infrastructures, thermal modelling, reliability, programming models, and timing analysis.

\subsection{Runtime resource management and reliability}

The problem of unexpected hardware failures is rapidly growing,
due to the increasing thermal stress sensitivity of hardware components in modern HPC systems.
For some applications, reliability issues may become critical,
because faults lead to silent data corruptions passing undetected by the hardware,
or otherwise because mere re-execution upon a crash may lead to violating the expected execution time.
In other words, for time-aware applications, the occurrence of some runtime error may lead
to deadline misses (timing violations), even when appropriate error recovery mechanisms, e.g. checkpointing, are put in place.
In the latest years, several \emph{Dynamic Reliability Management} solutions have been introduced~\cite{Farm2011,Haghbayan2016,Huang2010,Warm2016}.
Most operate on a \emph{reactive} basis, by exploiting checkpoint-restore protocols, to resume corrupted data or faulty application executions.
\emph{Proactive} approaches, based on suitable predictive models, have been recently proposed.
However, such models are often platform-specific and based on single-node computing systems.
Furthermore, they do not usually consider the application timing requirements.
The goal is typically to maximize the lifetime of the hardware components.

\subsection{Programming models}

HPC applications are usually programmed by using \emph{hybrid} programming models,
combining a distributed memory model, often MPI, to manage the application parallelism at global level,
and a shared memory model, to manage parallelism at the single-node level.
In heterogeneous systems, the node-level programming model needs specialized support for heterogeneity, such as those provided by OpenCL or CUDA. 
In such heterogeneous programming models, the programmer is in charge of managing explicitly the required resources,
which demands a large amount of boilerplate code.
When resource management is available, it is generally placed at the top level (in the application dispatcher)
or it operates under the assumption that an entire node is allocated to a single application.
Such coarse approaches pose an inherent limitation, that will be addressed by RECIPE to enable application-driven management strategies.

\subsection{Heterogeneity}
Current trends in HPC are increasingly moving towards heterogeneous platforms,
i.e. systems made of different computational units, ranging from general-purpose processors
to graphics processing units (GPUs) and even special-purpose units made of custom acceleration logic,
often implemented on field-programmable gate arrays (FPGAs)~\cite{bib:euroexa,bib:intelstratix,bib:mango,cilardo15}.
In fact, several players have recently introduced FPGA-based heterogeneous platforms
used in a large range of high-performance computing applications~\cite{paranjape2012heterogeneous},
e.g. multimedia, machine learning, bioinformatics, etc.~\cite{sarkar10,paranjape2012heterogeneous}, with speedups in the range of 10x to 100x.
However, the effective exploitation of combined heterogeneous resources still poses an open issue in current HPC,
and has only been demonstrated for very specific workloads.

\subsection{HPC interconnect}
Current HPC interconnect technologies include Ethernet~\cite{eth2017}, InfiniBand~\cite{ib2001}, as well as vendor specific interconnects, particularly Intel Omni-Path technology~\cite{op2015}.
Ethernet, a dominant standard for mainstream commercial computing requirements,
has continued to evolve, reaching performance levels of 400 Gbps in 2017.
InfiniBand is designed for scalability, relying on a switched fabric network topology together with remote direct memory access (RDMA) to reduce CPU overhead which enables maintaining a performance and latency edge in comparison to Ethernet in many high performance workloads.
The InfiniBand roadmap~\cite{bib:ibta} details bandwidths reaching 600Gb/s data rate HDR in the middle of 2018 and 1.2Tb/s data rate NDR in 2020.
Introduced in 2015, Intel's end-to-end Omni-Path Architecture (OPA) claims higher messaging rates
and lower latency than InfiniBand, in addition to advanced features such as traffic flow optimization, packet integrity protection and dynamic lane scaling,
although it is not currently a recognized standard.
In RECIPE, InfiniBand will be the HPC interconnect of choice, because of the higher performance versus Ethernet as well as its role as a de-facto standard, unlike Intel OPA.
However, the project will need to address crucial issues related to the interplay with the deeply heterogeneous acceleration fabric,
for which no solutions exist exposing direct access to HPC interconnect technologies,
as well as the interplay between the QoS mechanisms provided at RTMS and interconnect level.

\subsection{Thermal modelling}
Efficient thermal management requires accurate knowledge about the thermal profile of the chip in both steady and transient states.
In conventional Multi-Processor Systems on Chip (MPSoCs), the knowledge of the chip floorplan allows the design-time identification of the hot-spot locations.
However, this approach is no longer possible in heterogeneous MPSoCs equipped with a reconfigurable fabric,
as the thermal distribution depends on the accelerators implemented, which are unknown during the design phase.
As a result, energy efficiency comes at the cost of shifting thermal evaluation
from the chip design phase to the run-time management, affecting also runtime reliability \cite{zoni2019alldigital}.
This may be particularly relevant to RECIPE, because of the adoption of FPGA-based accelerators.
The project will thus aim at leveraging accurate, fast, and flexible thermal simulation
to demonstrate novel efficient thermal management strategies for heterogeneous HPC platforms.

% Efficient thermal management requires accurate knowledge about the thermal profile of the chip in both steady and transient states.
% In conventional Multi-Processor Systems on Chip (MPSoCs){\color{red} CH: I don't like the term MPSoC in the context of HPC,
% Also I would not to expect to have heterogeneity within the MPSoCs or chip but heterogeneous components/chips},
% the knowledge of the chip floorplan allows the design-time identification of the hot-spot locations.
% However, this approach is no longer possible in heterogeneous MPSoCs equipped with a reconfigurable fabric,
% as the thermal distribution depends on the accelerators implemented, which are unknown during the design phase.
% As a result, energy efficiency comes at the cost of shifting thermal evaluation
% from the chip design phase to the run-time management, affecting also runtime reliability.
% This may be particularly relevant to RECIPE, because of the adoption of FPGA-based accelerators{\color{red} CH: I don't follow this}.
% The project will thus aim at leveraging accurate, fast, and flexible thermal simulation
% to demonstrate novel efficient thermal management strategies for heterogeneous HPC platforms.

\subsection{Timing analysis}
The large amount of computation and storage nodes integrated in HPC systems makes applications 
suffer from considerable performance variability, which will inevitably increase with the advent of exascale HPC systems.
The uncertainty in the timing behavior of HPC applications is a consequence of either the internals of the application (intrinsic variability)
or the interactions between different applications or with the system itself (extrinsic variability).
Timing variability makes it difficult to achieve tight execution time bounds for the applications running in the computing infrastructure
\cite{Reghenzani2019pWCET}.
For applications with tight real-time requirements, like those addressed by RECIPE, standard practices of performance analysis are not enough
and new methods are required to derive trustworthy and tight upperbounds to application execution times 
\cite{reghenzani2018chronovise,8585132,EVTECRTS,MBPTACV}

\section{Architectural Concept}
\label{architectural-sec}
The RECIPE hardware architecture builds on the outcome of the MANGO H2020 project~\cite{flich2018mango}.
While the host-side part of the prototype includes commercial architectures
like Intel Xeon~\cite{bib:trinity, bib:marenostrum} as well as GPUs~\cite{bib:titan},
heterogeneous acceleration relies on the MANGO Field Programmable Gate Array (FPGA) fabric~\cite{bib:mango} in addition to commercial HPC FPGA cards.
The MANGO prototype, originally aimed at manycore architecture exploration,
contains two main components, the General-purpose Node (GN), which includes commercial CPUs and GPUs, and the Heterogeneous Node (HN),
made of customized motherboards mounting dedicated FPGA daughterboards developed by ProDesign GmbH.
HNs offer a large degree of flexibility for exploring and instantiating customized accelerators and various heterogeneous system configurations.
Among other features, the key architectural aspects of MANGO inherited in RECIPE include
the communication infrastructure between the GN and HN,
synchronization mechanisms between the host and the hardware application kernels,
QoS guarantees, burst transfers, unified HN memory address space, and a basic HN hardware abstraction library.

An important challenge in RECIPE relates to the increased scale of the target HPC system.
While MANGO focuses on the concept of an HN node made of several interconnected FPGAs,
RECIPE targets a more ambitious scenario where several HN nodes coexist in the system, each defining its own set of accelerators.
While the HNs are physically distributed throughout the HPC system, they are part of a single pool of accelerators.
As an example, in MANGO an application can obtain a set of resources, like memory or accelerators, only from its corresponding HN node.
In RECIPE we extend the application capabilities to use accelerators from different, efficiently interconnected HN nodes at the same time.

In order to fully match the potential of the RTMS, the RECIPE heterogeneous accelerators
also expose hardware monitors and a suitable hardware abstraction layer, described in Section~\ref{hal-sec}.
The hardware abstraction layer provides a uniform view of the available accelerators, including GPU, manycore, and FPGA devices,
exposing fine-grain partitioning mechanisms used to ensure scalability, system-wide reconfigurability, and dedicated resource provisioning. 
%
% In that respect, the project will build on the results of the MANGO FET-HPC project~\cite{flich2016mango,flich2015mango,Flich2018},
% both in terms of deeply heterogeneous accelerator architecture and of resource management framework.

% The project will also need to address important challenges related to the integration of large-scale reconfigurable hardware in user applications.
% In particular, for the development of its heterogeneous abstraction layer, RECIPE will explore four different styles of use for FPGA acceleration,
% depending on the specific use case requirements:
% \begin{itemize}
% \item full custom HDL implementation, suitable for performance-critical, relatively simple and regular kernels;
% \item optimized library-based design, for well-supported kernels to be implemented in hardware, e.g. linear algebra;
% \item pure-hardware HLS-based design, for non-standard kernels or control-intensive parts of the application that are not performance-critical;
% \item software-programmed accelerators, particularly the NaplesPU vector core~\cite{NPU} and associated LLVM-based compiler imported from MANGO, which is suitable for 
% control-intensive parts of the software application that do not match the restrictions of HLS and/or data-intensive kernels benefitting from custom vector-style 
% approaches.
% \end{itemize}

%\subsection{Interconnect configurability}
Similar to the multi-accelerator compute platform, the memory/interconnect architecture also needs to scale up to the system level.
The MANGO solution allows some basic bandwidth allocation strategies {\em within} the HN,
in that HN accelerators and the connected server can obtain reserved interconnect bandwidth and memory resources,
although this feature is limited to the applications running on the particular server connected to the HN.
These capabilities are expanded in RECIPE from the node level to the system level,
totally decoupling memory and interconnect management from the node boundary and exposing such functionality to the complete system.
In order to achieve that purpose, we augment the MANGO prototype with a standard InfiniBand adapter, plugged into the GN part of the MANGO cluster,
which in turn enables the integration of the reconfigurable accelerator fabric within the HN in the RECIPE prototype,
such that the FPGAs inside of the HN can be seen as other heterogeneous devices in the system.
Unlike QoS, reliability-related mechanisms are not directly supported by the MANGO technology.
However, MANGO allows multiple accelerator {\em replicas} to be instantiated and used on the same HN or on different HNs.
This feature paves the way to new flexible mechanisms for enhancing fault-tolerance in RECIPE.
In fact, in the project we provide node-level (within the HN node) and system-level (within the Resource Manager) reliability mechanisms 
able to detect and isolate malfunctioning devices located in the HN node, then exploiting
the underlying reconfigurability to enforce proactive and reactive fault-tolerance strategies.

\begin{figure}[ht!]
	\centering
	\includegraphics[width=0.8\linewidth]{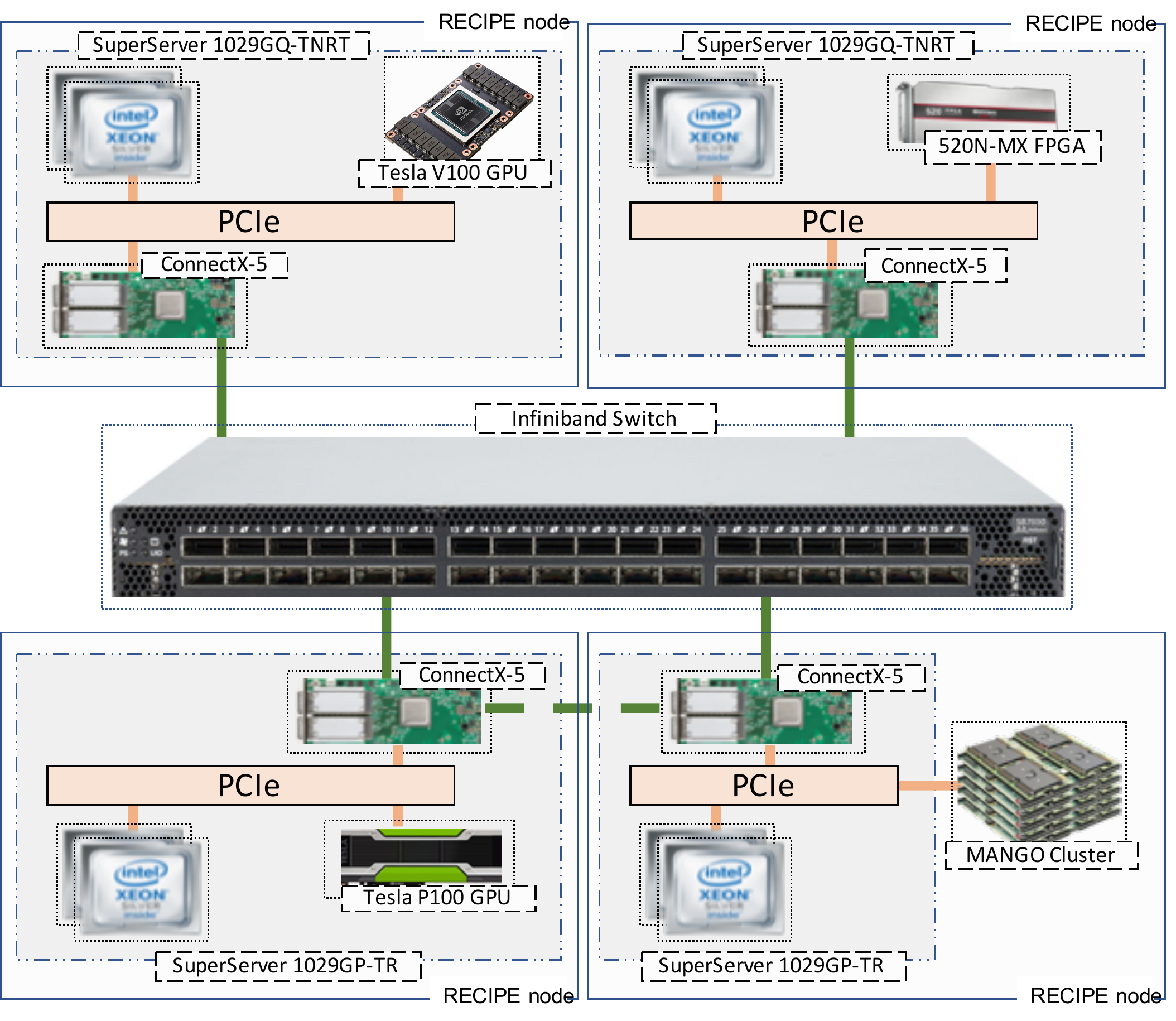}
	\caption{RECIPE prototype}
	\label{fig:proto}
\end{figure}

\subsection{The RECIPE prototype}

Figure~\ref{fig:proto} illustrates the prototype built in the context of RECIPE. It is composed of a set of heterogeneous nodes interconnected through a high-performance InfiniBand network. The prototype includes commercial CPUs and  GPUs, plus commercial of-the-shelf (COTS) FPGAs and the MANGO HN cluster. This assembly results in a deep heterogeneous system that incorporates the most common state-of-the-art heterogeneous resources for HPC systems.

\subsection{Hardware abstraction layer and low-level runtime support}
\label{hal-sec}
One of the challenges that need to be addressed when designing highly heterogeneous systems lies on keeping the programability of such systems under control, meanwhile diminishing any sort of performance degradation. In fact, different compute architectures mean distinct instruction sets (ISAs) and/or programming flows, thus making the programming of these systems complex. In RECIPE, we rely on the a local resource manager, a hardware abstraction layer (HAL) and a custom low-level runtime environment to deal with this demanding task. 

The HAL developed for the RECIPE project provides a unique C-based programming application interface to: I) access the different type of heterogeneous resources targeted, II) support the different acceleration modes supported by RECIPE heterogeneous accelerators, and III) expose to the applications and the to the local resource manager, a global and unified view of these resources, which can be accessed independently of the physical location they are located, mimicking the concept of resource dissagregation already explored in datacenters for both CPU and GPU devices.

\begin{figure*}
	\centering
	\includegraphics[width=0.8\textwidth]{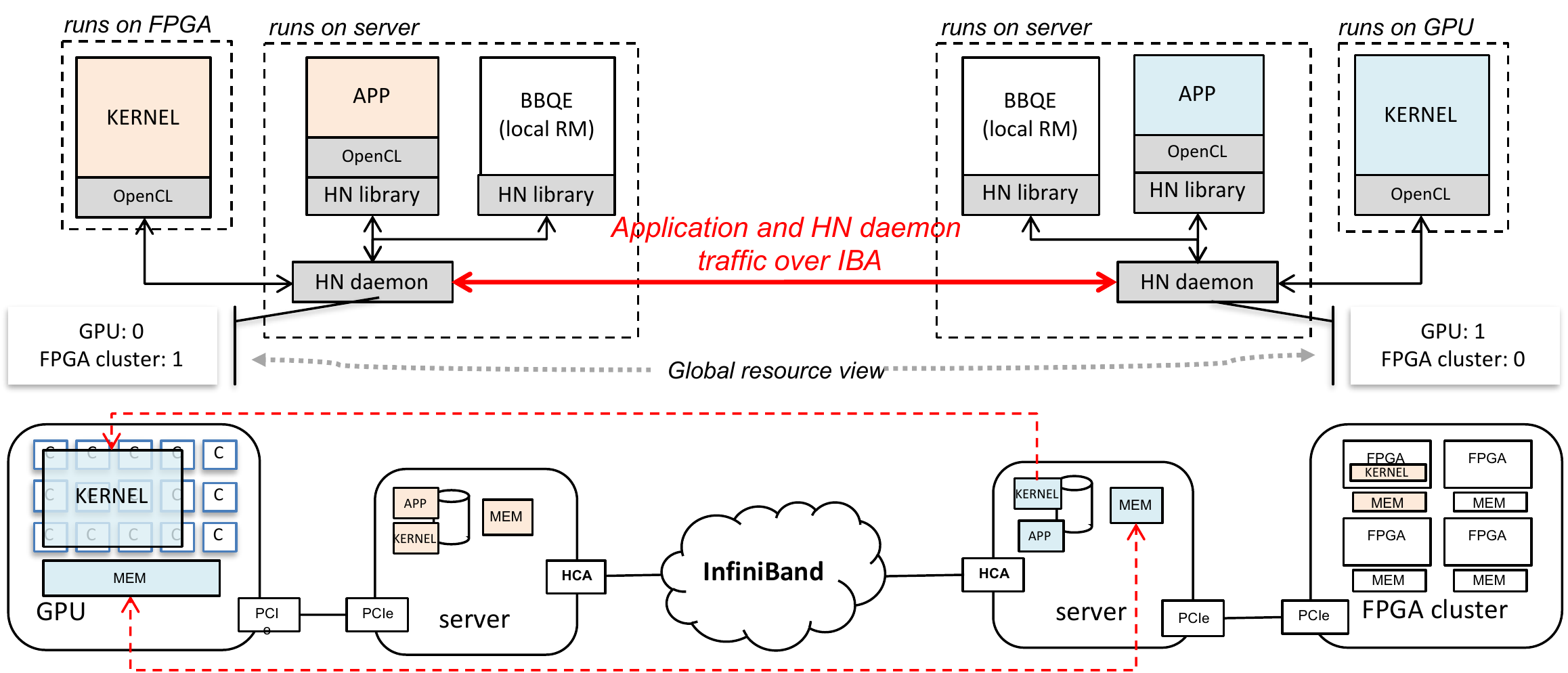}
	\caption{Low-level runtime architecture.}
	\label{fig:runtime}
\end{figure*}

To satisfy its goals, the HAL relies on the distributed low-level runtime environment that is shown on Figure~\ref{fig:runtime}. It consists of two main components: the HN library and the HN daemon.  On the one hand, the HN library is the part that an application process, that wants to access the RECIPE heterogeneous resources, has to be linked to. This library includes monitoring, synchronization and communication support, among other features. In general, it establishes communication with the HN daemon for: obtaining the different resources the system is composed of, together with status information of them; setting QoS application requirements; allocating memory and computing resources; transferring data and launching kernels. 

The HN daemon is a distributed user space process that captures the HN library requests and it executes them on the correct accelerator or resource device. As a distributed process, there is an instance of it running in each system node. These instances discover themselves by means of a hierarchical discovery service which allows every instance to have the same global view of resources, plus the communication cost for accessing each device, that depends on the node that is used to access them. In the figure, this process is illustrated with the \emph{box calls} connected to the HN daemon. As it can be seen, both processes can manage the same heterogeneous resources, but the cost to access them differs depending on the number of network hops required to reach them. In that way, the daemon on the left of the figure has direct access to the GPU, but a cost of one hop to access the remote FPGA cluster, since we are assuming that the servers in the figure are exactly one host distance away each other for this example. Thus, the communication cost is a useful value that should be taking into account for the scheduling and mapping strategy implemented by the resource manager.

To offer the low-level runtime services to the applications through the HAL in a simple manner the HN library provides a library wrapper to capture OpenCL function calls, so that OpenCL-based applications can run in the system without modifications. In addition, these applications can access every resource in the system in a location-independent way. This is shown in the figure with the dash red lines. As it can be seen, an OpenCL-based application (marked in blue color) running on the server of the right is using the remote GPU to execute its OpenCL kernel.

%The traffic generated by the applications and the HN daemons goes through an InfiniBand network, which will allow to the resource manager to change the QoS policies dynamically according to its needs. In addition,

\section{Runtime management system}
\label{rtms-sec}

RECIPE aims at addressing the reliability problem by operating both at the node
and at the infrastructure level, thanks to a {\em hierarchical} resource
management approach.  Our objective is to integrate reactive and proactive
solutions, while maximizing the lifetime of the system, as well as providing
reliability guarantees to time-sensitive applications.  For the latter, we need
to minimize the use of reactive mechanisms and maximize the effectiveness of
predictive resource management, thus supporting the combined optimization of
reliability and timing.

Based on this approach, the RECIPE runtime management system will be
partitioned in two layers, a {\em global} resource manager providing proactive
fault tolerance and coarse-grained resource allocation, as well as a {\em
local} resource manager, ensuring reactive fault tolerance and fine-grained
resource allocation.
In particular, RECIPE will build on the integration of BarbequeRTRM
\cite{Bellasi2015Effective} and the SLURM-based Global Resource Management
(GRM) framework.  These will cooperate on a hierarchical basis, with GRM being
in charge of dispatching the workload at the global infrastructure level, and
BarbequeRTRM mapping the application's tasks and memory requests at the local
node level.
Each instance of BarbequeRTRM (one per node) will take into account the timing
requirements of the application dispatched by the GRM layer, the current status
of the node, i.e. load and power consumption when available, and the
performance profile of the available (heterogeneous) computing resources.
% The resource allocation policy tries to perform resource assignments
% that could guarantee the required performance levels, minimizing the system power consumption.
To this aim, the BarbequeRTRM relies on the hardware abstraction layer
described in Section \ref{hal-sec}, to get knowledge of the heterogeneous set
of processing units and memory nodes installed into the system, and enforce the
resource allocation decisions of the management policies. The specific case of
processing units of type CPU instead, is based on the abstraction and the control
capabilities offered by the Linux frameworks \emph{cgroups} and \emph{cpufreq},
to isolate the ``host-side" application threads and perform frequency scaling.

As specific technical actions planned by RECIPE, we will extend the BarbequeRTRM capabilities by:
\begin{inparaenum}
\item feeding the policies with data coming from suitable models, in order to
implement \emph{reliability-aware} and \emph{timing-driven} resource allocation policies;
\item providing a local checkpoint-restore mechanism, transparent with respect to the
application, to resume the execution of application's tasks and optionally redefine
the set of assigned resources.
\end{inparaenum}
On the other hand, the GRM capabilities will also be extended by:
\begin{inparaenum}
\item adding a proactive reliability module that will communicate with the
3D-ICE~\cite{3d-ice} temperature modelling module, enabling reliability
modelling and prediction; 
\item proposing new reliability-aware runtime global management strategies; and
\item enhancing the capabilities of distributed computing of the GRM, enabling
the application execution to span across multiple nodes.  
\end{inparaenum}

In order to support the above predictive strategies, RECIPE will develop novel
models integrated with run-time management policies, running and cooperating at
both the node and the infrastructure level. Furthermore, to assess the impact
of reliability and thermal issues on the overall system architecture, we will
also rely on simulation, which will be possible thanks to the recent gem5-X
framework. In RECIPE, gem5-X~\cite{gem5X} will be equipped with both thermal
and reliability simulation to evaluate the impact of architectural choices and
resource management strategies in short and long term reliability.  We will
investigate state-of-the-art techniques in fault prediction
technologies~\cite{faultprediction}.  Existing fault-prediction techniques are
generally based on the existence of a model of the system that has the ability
to predict the occurrence of faults when the real-time performance, power, and
temperature measurements of the different system components indicate that
faulty situations are likely to occur.
These models can be built by using fault statistics of systems based on similar
technologies, when available, or by using information provided by the component
manufacturer which allows creating simulation models of the system.  When both
approaches are combined, synthetic models can be enriched or calibrated at
runtime with information collected during the system lifetime.  Given that the
system temperature is a very important contributor to the component
reliability, fault prediction models will also incorporate the thermal
behavior.
The fault/performance prediction, based on application statistics, will then
drive the runtime resource manager during scheduling and task mapping, in order
to minimize missed deadlines and adapt the checkpoint rates to various fault
recovery strategies, on a given HPC system configuration. In the experimental
section, we carried out a preliminary analysis of three first possible policies
for tuning the checkpoint rate, observing the expected impact on the
application execution time.

\subsection{Reliability}

A reliable HPC infrastructure must include \emph{Checkpoint/Restore (C/R)}
mechanisms, in order to enable the possibility of recovering the execution of
applications experiencing faults at run-time. This is a typical \emph{reactive}
approach to fault recovery. In RECIPE, we aim at minimizing the probability of
triggering reactive actions, due to the costs of the restore procedures.
However, our runtime management software also implements these mechanisms in order
to improve system robustness by covering cases of misprediction of
the reliability models.
Checkpoint/Restore techniques are normally classified in application-level,
user-level, and system-level solutions~\cite{Egwutuoha2013}. The latter is
operated by the system hosting the job in a transparent way and in turn is divided
in solutions relying on Operating System support and hardware-level support.
We base our C/R technique on a system-level approach, inherently needing a
combined support from the underlying hardware and OS/RTMS.
This means that, from the programming model perspective, the involvement of the
application developer is not required. Checkpoints are periodically performed,
by the resource managers, which are therefore in charge of handling the 
trade-off between reliability and overhead, according to the application and
the overall system requirements.

In addition, as one of its main objectives, RECIPE directly addresses the heterogeneous nature of modern HPC systems,
particularly \emph{deeply} heterogeneous architectures based on FPGAs.
In that respect, given the importance of Checkpoint/Restore in traditional HPC, RECIPE aims at extending
the C/R approach to FPGA accelerators, providing a new reliability mechanism for emerging classes of HPC applications.
We highlight that the proposed mechanism is not alternative to the range of low-level techniques for FPGA reliability~\cite{Lee2017,Cheatham2006,Parris2011},
which can coexist with the RECIPE solution since they involve either the physical device or the particular user design.
Rather, the FPGA-extended C/R techniques explored in RECIPE provide a way of effectively controlling the FPGA-side execution state from the host system,
so as to checkpoint and restore it as part of a larger application state.
As such, the proposed approach is orthogonal to the type of faults, e.g. aging or single event upsets,
which might affect the FPGA device and, indeed, it is instrumental to preserving the integrity of the system-wide application state
upon faults involving any part of the HPC infrastructure.

More specifically, the FPGA C/R techniques explored in RECIPE can either take a coarse-grain
approach, where the whole state of the acceleration card (the FPGA device chip as well as the off-chip memory) is saved and restored, or a
fine-grain approach, where a \emph{permanent state} can optionally be defined
by the user, indicating selectively what subset of the overall physical state
is of actual relevance for checkpointing, involving both the FPGA device and
the off-chip memory. For \emph{off-chip} state, i.e. DDR or HBM memory
available in the card, the user is required to explicitly indicate the memory
address windows that contain the permanent state, while for \emph{on-chip} state, the user is required to instantiate in the FPGA design special library
components, provided by RECIPE, featuring stateful behavior (registers,
configurable memory modules, etc.). The explicit use of such components tells
the RECIPE flow that the information held by the corresponding instantiated
modules is part of the permanent state.
Regarding the access mechanism, RECIPE considered a range of different solutions,
whose actual availability might depend on the specific FPGA device. 
In particular, the checkpointing and restoring of the FPGA execution state may be supported through dynamic partial reconfiguration by means of internal ports,
such as Xilinx Internal Configuration Access Port (ICAP), which provide the ability of reconfiguring the device from the design itself, requiring however a
suitable instrumentation of the hardware design.
Alternatively, device state readback and restore operations, in either a global or a partial form, can take place through JTAG, a well-established
industry standard for verifying and debugging integrated circuits and printed
circuit boards, which all FPGA devices normally provide in order to upload the
configuration image to the device as well as retrieve data from the device. In the case of the JTAG access mechanism, for
accessing the off-chip memory one must either require that the external memory
device (a DDR device or a HBM chiplet) has itself a JTAG port accessible on the
same scan chain or, as an alternative solution, make the FPGA-side memory
controller accessible from the JTAG logic and, hence, from the host-side
software. In the Xilinx devices used in RECIPE, for example, this is possible
by means of a special JTAG-to-AXI master component, which essentially allows
the FPGA JTAG port to control an additional AXI master in the FPGA, competing with the user
design for access to the memory controller and hence to the off-chip memory.

\subsection{Interplay with programming models}

As highlighted above, most RTMS-level choices will be driven by
application-specific QoS requirements and behavior profiles.  This is
particularly important because RECIPE envisions scenarios where multiple
unrelated applications will co-exist on the same nodes, thus requiring
system-wide resource management in order to maximize utilization of (possibly
heterogeneous) compute resources.
To support resource management at the node level, the application needs to be
able to interact with the resource manager in a way that is as much transparent
as possible to the programmer, so as to provide the resource manager with the
appropriate information about the required Quality of Service and the type of
resources that the application can effectively exploit.
The interface between the runtime management system and the application in
RECIPE will be mediated by the programming model, supporting the expression of
such application expectations.
Existing experimental models, however, do not fully match the requirements of
typical industry-grade implementations.  In particular, there is not support
for dynamic compilation, nor for any source language except C/C++. 
In RECIPE, we will build on the \emph{mangolib} model and implementation
developed in the MANGO FETHPC project~\cite{flich2018mango} to integrate
system-wide resource management within the programming model, and will extend
it to reach industry-grade maturity.
The current \emph{mangolib} model allows the application to specify a different
``recipe'' for each kernel, as well as multiple implementations of the same
kernel, targeting different accelerators.  The runtime manager then
automatically manages the selection of the best configuration according to the
application priority level, quality of service requirements, and resource
availability, while optimizing system-wide metrics of utilization and
energy-efficiency.
%
%We will further extend the model to support multiple languages, depending on
%application requirements.  Candidate extensions include improved compatibility
%with the OpenCL programming model and the FORTRAN programming language
%\cite{massari2014extending}.  Furthermore, we will provide dynamic compilation
%capabilities to allow just-in-time compilation of kernels from the source
%code.

In order to minimize the effort in terms of application porting, we recently
built an OpenCL wrapper layer on top of \emph{mangolib}.  This layer comes with
a very minimal set of changes in the common OpenCL API specification. In
Listing \ref{lst:opencl-mango}, we show a classical example of OpenCL sample
application: \emph{matrix multiplication}, implemented by using the custom
OpenCL API.  The functions introduced or modified have been highlighted in red.
Here below, we can summarize the porting steps that a developer should carry
out to adapt an OpenCL standard compliant implementation to the run-time
managed version:

\begin{enumerate}
  \item Replace the original OpenCL header file \texttt{CL/cl.h}, with the \texttt{mango\_cl.h} header, coming with the
      \emph{mangolib} installation (line 1).
  \item Instantiate a \texttt{mango\_data} data structure, and fill it with information regarding the application name and
      the ``recipe'' file name, including the per-kernel requirements (lines 17-19).
  \item Pass the \texttt{mango\_data} structure to the \texttt{clCreateCommandQueue} function as 5th argument (line 20).
  \item Once the buffers have been created and populated (lines 53-58), pass them to the \texttt{clEnqueueNDRangeKernel},
      in order to bind the kernel to their input and output buffers (lines 79-83).
  \item Call the \texttt{clStartComputation} to trigger the resource mapping and, afterwards, the kernel execution (line 88).
\end{enumerate}

\definecolor{mygray}{rgb}{0.6,0.6,0.6}
\lstset{
  basicstyle=\scriptsize\ttfamily,
  xleftmargin=\parindent,
  language=C,
  frame=single,
  numbers=left,
  numberstyle=\tiny,
  commentstyle=\color{mygray},
  stringstyle=\it,
  keywordstyle=\bf\color{red},
  morekeywords={clStartComputation, clEnqueueNDRangeKernel, clCreateContext, mango_data},
  deletekeywords={context, int, sizeof, struct, void, \#ifdef, \#else, \#endif, char, return, const, unsigned},
  label={lst:opencl-mango},
  caption={Matrix multiplication: OpenCL implementation on top of \emph{mangolib}}
}

\begin{lstlisting}
#include <mango/mango_cl.h>

int main()
{
   int err;

   // Platform and device retrieval are kept for code portability
   cl_uint dev_cnt = 0;
   clGetPlatformIDs(0, 0, &dev_cnt);
   cl_platform_id platform_ids[10];
   clGetPlatformIDs(dev_cnt, platform_ids, NULL);

   cl_device_id device_id;
   clGetDeviceIDs(platform_ids[0],
		CL_DEVICE_TYPE_ALL, 1, &device_id, NULL);

   // Add MANGO library specific data to the OpenCL context
   struct cl_mango_data mango_data;
   mango_data.application_name = "matrix_multiplication";
   mango_data.recipe           = "mm";
   cl_context context = clCreateContext(0, 1,
				&device_id,
				NULL,
				&mango_data,
				&err);

   cl_command_queue commands = clCreateCommandQueue(context,
				device_id,
				0,
				&err);
   cl_program program;
#ifdef BUILDKERNEL
   char *KernelSource;
   // Load the kernel source for online compilation
   program = clCreateProgramWithSource(context,
 		1, (const char **) &KernelSource, NULL, &err);
#else
   char* KernelBinaryPath = "yourpathtobin";
   program = clCreateProgramWithBinary(context,
			1,
			&device_id,
			NULL,
			(const unsigned char **) & KernelBinaryPath,
			NULL,
			&err);
#endif

   err = clBuildProgram(program, 1, &device_id, NULL, NULL, NULL);

   // Creation of the OpenCL kernel object
   cl_kernel kernel = clCreateKernel(program, "matriMul", &err);

   // Host-side allocation of buffers for the matrices
   int* h_A = (int*) malloc(mem_size_A);
   int* h_A = (int*) malloc(mem_size_B);
   int* h_C = (int*) malloc(mem_size_C);

   // Creation of the OpenCL buffer objects
   cl_mem d_A = clCreateBuffer(context,
			CL_MEM_READ_WRITE | CL_MEM_COPY_HOST_PTR,
			mem_size_A,
			h_A,
			&err);
   cl_mem d_B = clCreateBuffer(context,
			CL_MEM_READ_WRITE | CL_MEM_COPY_HOST_PTR,
			mem_size_B,
			h_B,
			&err);
   cl_mem d_C = clCreateBuffer(context,
			CL_MEM_READ_WRITE,
			mem_size_C,
			NULL,
			&err);

   // Kernel arguments
   cl_mem in_buffers[2], out_buffers[1];
   in_buffers[0]  = d_A;
   in_buffers[1]  = d_B;
   out_buffers[0] = d_C;
   int wA = NR_ROWS_MATRIX_A;
   int wC = NR_ROWS_MATRIX_C;
   err = clSetKernelArg(kernel,  0, sizeof(cl_mem), &d_B);
   err |= clSetKernelArg(kernel, 1, sizeof(cl_mem), &d_A);
   err |= clSetKernelArg(kernel, 2, sizeof(cl_mem), &d_C);
   err |= clSetKernelArg(kernel, 3, sizeof(int),    &wA);
   err |= clSetKernelArg(kernel, 4, sizeof(int),    &wC);

   // Put kernels and buffers together
   err = clEnqueueNDRangeKernel(commands,
			kernel,
			&in_buffers, 2,
			&out_buffers, 1,
			0,
			NULL, NULL);

   // Asynchronous wait for the kernel to complete the
   // writing of the output buffer
   err = clEnqueueReadBuffer(commands,
			d_C,
			CL_TRUE,
			0,
			mem_size_C,
			h_C,
			0,
			NULL, NULL);

   // MANGO-specific application execution trigger
   clStartComputation(commands);

   free(h_A);
   free(h_B);
   free(h_C);
   clReleaseMemObject(d_A);
   clReleaseMemObject(d_C);
   clReleaseMemObject(d_B);
   clReleaseProgram(program);
   clReleaseKernel(kernel);
   clReleaseCommandQueue(commands);
   clReleaseContext(context);

   return 0;
}

\end{lstlisting}
  
With this implementation, the kernel-to-device mapping is tranparently managed
by the local resource manager, as mentioned above. This
means that the platform and device selection calls (lines 7-15) do not have any
real effect on the kernel compilation and offloading, until the function
\texttt{clStartComputation} is called. Here the programming library waits for
the resource manager to provide all the mapping information, before launching
the actual kernel execution. The remaining part of the application (host-side)
code is perfectly OpenCL standard compliant.

\subsection{Proactive/reactive thermal management}

To enable effective thermal management, RECIPE develops a holistic simulation
tool for a wide range of cooling techniques and settings, building on 3D-ICE
and using the most recent pluggable heatsink models that enable simulation of
arbitrary cooling models~\cite{heatsinkplug}, and using the gem5-X framework
\cite{zoni2015modeling}. The goal of these tools is to model and simulate
temperature in a spatio-temporal way, feeding the data to the reliability
models to estimate MTTF. 

For this to be a reality, the advances and enhancements undertaken in RECIPE
are as follows: 
\begin{itemize}
\item Enabling the simulation of arbitrary state-of-the-art cooling mechanisms,
such as the ones found in current servers. In particular, within RECIPE we need
to enable the simulation of both natural convection mechanisms (i.e.,
heatsinks), and forced convection cooling (i.e., heatsinks plus fans). This is
done by incorporating a pluggable heatsink model into the 3D-ICE thermal
simulator, which we have released as version 2.2.7 of 3D-ICE.
\item Proposal of a methodology that will allow us to assess the impact of the
main control knobs related to temperature in today's servers, which range from
workload allocation, DVFS setting and fan speed control
policies~\cite{IranfarDATE2020}. 
\item Proposal of a methodology to adequately link the thermal aspects to the
reliability of the system. For this purpose, we will use the MTTF and
reliability models proposed, which will be incorporated into the resource
management policies.
\end{itemize}

To accomplish this goal, we need ways of validating temperature against real
devices. For this purpose, within the RECIPE project we have also created a
platform which comprises a real test chip~\cite{terraneo2019open} for accurate
thermal characterization. In particular, this platform is based on a Thermal
Test Chip (TTC), an integrated circuit containing an array of power dissipating
elements and an array of temperature sensors. Our thermal platform is capable
of applying a generic power dissipation pattern to the thermal test chip and
measuring the corresponding temperature map, at a rate up to 1 kHz. This
capability allows to measure the temperature map of an integrated circuit
subject to reference power dissipation maps, and thus to design and validate
thermal models. In our particular case, the chip is organised as a 4 by 4 array
of individual cells, each capable of temperature sensing and power generation
through a resistive element. The heating element in each cell is capable of
dissipating up to 12 W. This allows to simulate the equivalent of a 16-core
chip with a total power dissipation of 192 W. 

\begin{figure}
\centering
\includegraphics[width=0.8\columnwidth]{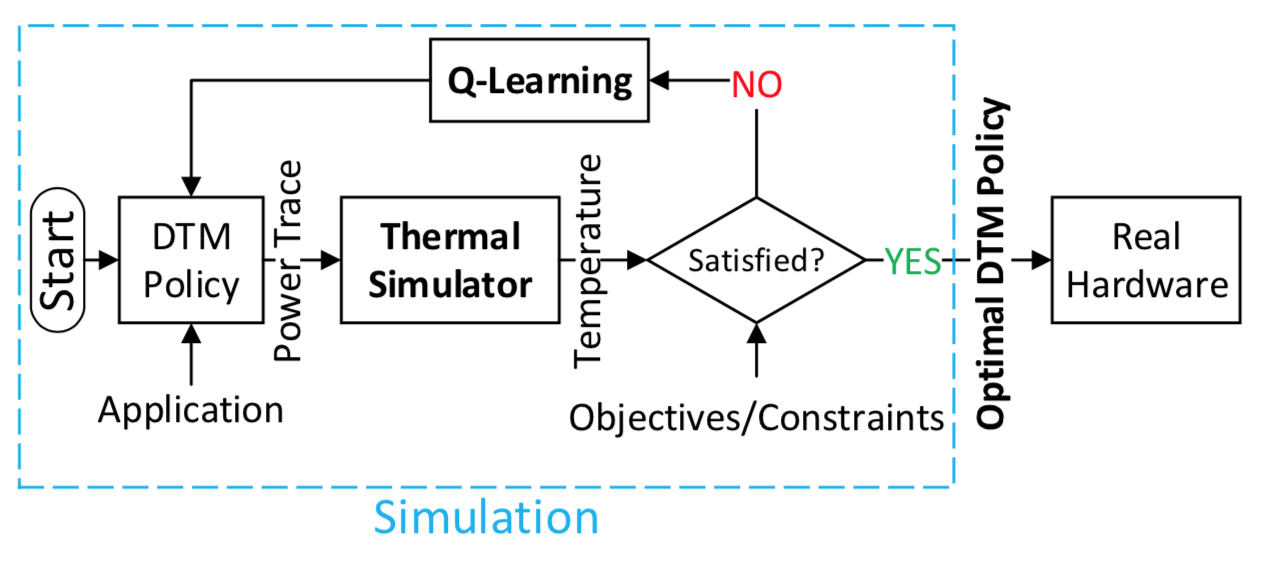}
\caption{Thermal modelling methodology to enable DTM and reliability management~\cite{IranfarDATE2020}}
\label{fig:method-thermal}
\end{figure}

The methodology we envision for enabling DTM and reliability management is the
one depicted in Figure~\ref{fig:method-thermal} and described
in~\cite{IranfarDATE2020}. The simulator will enable us perform an exploration
of the impact of aggressive DTM management policies which may not only cause
performance degradation and additional power consumption, but more importantly,
jeopardizes lifetime reliability of the whole system. In fact, one of the main
reasons that makes researchers reluctant to consider fan speed control as a key
DTM approach is the lack of a transient thermal simulator for MPSoCs with
proper integration of fans. The incorporation of fan models in 3D-ICE, and the
use of this methodology provides a comprehensive framework for exploring
thermal effects of DTM policies in a safe way.

Workload allocation, DVFS and fan speed altogether drastically increase the
number of runtime design parameters to be decided by a DTM and
reliability-aware policy, which leads to additional challenges to find the best
values for optimal behavior of the whole system. Our methodology enables the
exploration of this design space in an automated way. In particular, we
envision the use of Reinforcement Learning (RL) techniques. The RL agents can
explore the design space using 3D-ICE and combine that with the impact of
reliability. Initial experiments~\cite{terraneo2019open}, prove the feasibility
of our approach.

\subsection{Timing constraints management}
%\subsection{Probabilistic worst-case application execution time computation}

\label{wcet-sec}
RECIPE has developed a methodology for HPC application analysis of timing
behavior for driving the decisions made by the RTMS. Our approach builds upon
techniques based on memory layout randomization for increasing test
coverage~\cite{Stabilizer,SWRAND}, as well as measurement-based probabilistic
timing analysis (MBPTA)~\cite{EVTECRTS,MBPTACV} to predict rare (high) timing
behavior beyond those values observed in applications with a continuous flow of
(big) data.

This means that the application can come with timing requirements in terms of
tasks completion times (deadlines) and related probability value \emph{p}.
Such a value represents the required probability of meeting the deadline
specified. In other terms, $1-p$ is the acceptable probability of experiencing
deadline misses. Accordingly, a suitable characterization of the timing
responses of the application must be performed, considering the heterogeneous
set of resource allocation options that the RTMS must take into account.  Then,
at run-time, the RTMS can exploit this characterization to drive the actual
process of resource mapping, thus selecting the combination of processing units
and the memory resources that should provide the required timing guarantees.

In this paper, we report the characterization of a geophysical
exploration application which builds upon Full Waveform Inversion (FWI), a
cutting edge technique that aims to acquire the physical properties of the
subsoil from a set of seismic measurements~\cite{fwi1,fwi2}.

The complete tool flow for probabilistic worst-case execution time provides:
\begin{enumerate}
\item Exploration of execution conditions. We integrate a software
randomization layer in the geophysical exploration application to test its
susceptibility to memory layouts caused by different code, heap and stack
allocations.% produced by the operating system.  
\item Worst-Case Execution Time (WCET) analysis. We analyze and fit an MBPTA
technique for WCET prediction so that it can be used in the context of HPC
applications running on high-performance systems.
\end{enumerate}

In the remaining of this section, we evaluate those techniques on the
geophysical exploration application, proving their viability to study its
(high) execution time behavior, and we show that appropriate integration of
those techniques allows scaling the application to the use of parallel
paradigms, thus beyond the execution conditions considered in embedded systems
-where WCET and pWCET estimates were originally proposed and used.

In particular, MBPTA has been applied at the application level. To measure the
accuracy of our approach, we compute the MET (\emph{Maximum Execution Time}) of
the application in 1000 independent runs. We also compute the MET when using
randomization to measure the accuracy of the pWCET for each of the
randomization techniques. The application was run in isolation in the system (a
common setup for critical or large applications in HPC infrastructures). More
details can be found in~\cite{Fusi2020}.

\begin{figure}[t!]
\centering
\includegraphics[width=0.8\columnwidth]{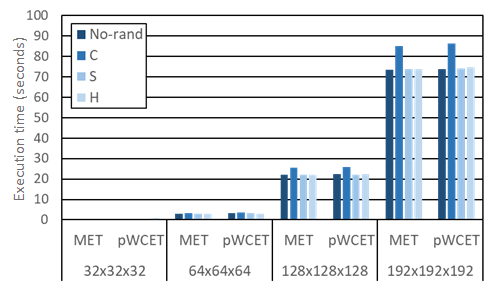}
\caption{Absolute maximum execution time (MET) and pWCET estimate (in seconds) at an exceedance probability of $10^{-6}$  for the different configurations of the application (FWI).}
\label{fig:METpWCETabs}
\end{figure}

\begin{figure}[t!]
\centering
\includegraphics[width=0.8\columnwidth]{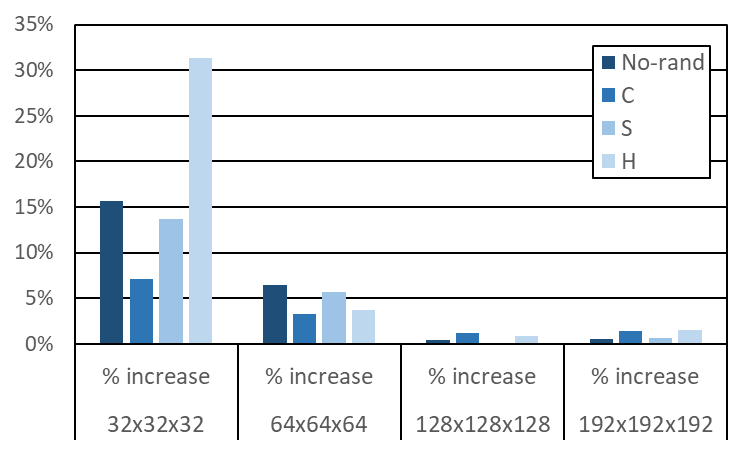}
\caption{Relative increase of the pWCET estimate at an exceedance probability of $10^{-6}$ w.r.t. maximum execution times (MET) for the different configurations.}
\label{fig:METpWCETrel}
\vspace{0.2cm}
\end{figure}

Finally, we show the absolute maximum execution time (MET) and pWCET estimates
for different problem sizes at an exceedance threshold of $10^{-6}$ per run in
Figure~\ref{fig:METpWCETabs}. We show results for -memory layout- randomization
of code (C), stack (S) and heap (H). We also show no-randomization for
comparison purposes. Overall, absolute differences are small, thus showing that
pWCET estimates are very close to the MET out of 1,000 runs in absolute terms,
which indicates that unobserved execution times can only be slightly higher
than observed ones. Moreover, results show that stack and heap randomization
cause little performance variation. However, code randomization causes a
noticeable impact in MET and pWCET estimates, thus showing that code memory
placement is particularly critical for this application.
Relative results on the increase of the pWCET estimates w.r.t. MET are shown in
Figure~\ref{fig:METpWCETrel}. As we increase problem size, execution time
variability decreases in relative terms, which leads to sharper pWCET
distributions and hence, smaller increase in relative terms of the pWCET
estimates w.r.t. MET. Hence, the larger the problem size is, the more stable
the execution time is. The use of SWrand \cite{SWRAND} along with MBPTA-CV
provides strong guarantees on these observations, thus allowing an efficient
use of resources scheduling HPC applications accounting for their unobserved
high execution times. Further analysis can be found in~\cite{Fusi2020}.

To the best of our knowledge, no specific methods have been devised to predict
reliably high execution times of HPC applications. Hence, our adaptation of a
mechanism –MBPTA-CV – used in another domain to fit the needs and constraints
of HPC systems is a pioneering attempt to reach the goal of reliable and tight
estimation of execution time bounds for HPC applications.

\subsection{Impact of prediction-based checkpointing policy}

In this subsection, we focus on how the run-time resource management
infrastructure can minimize the checkpoint overhead, which introduces some
penalty in the workload execution times, by exploiting the aforementioned
models. The goal is therefore, to minimize the checkpoint rate, controlled
by the local resource manager, according tothe expected failure probability.

To assess the impact of prediction-based checkpointing, we performed a
preliminary analysis, by comparing three possible policies.  As a baseline, we
considered a standard reactive \emph{restart} mechanism -- when the application
fails, the system restarts it. No checkpointing is needed, but for many use
cases the restart cost is not acceptable. With the periodical checkpointing
feature enabled instead, we considered different possible policies to control
the checkpoint rate:
\begin{itemize}
  \item \emph{fixed-rate}: periodical checkpoints of managed applications performed at regular intervals;
  \item \emph{prediction-based}: checkpoint the application in the immediate proximity of a predicted failure;
  \item \emph{error-tolerant}: considering the predicted MTTF, attempts to checkpoint at 90\% of the predicted MTTF,
  to account for the inherent inaccuracy of the prediction models.
\end{itemize}

\begin{figure}[t]
\centering
\includegraphics[width=0.95\textwidth]{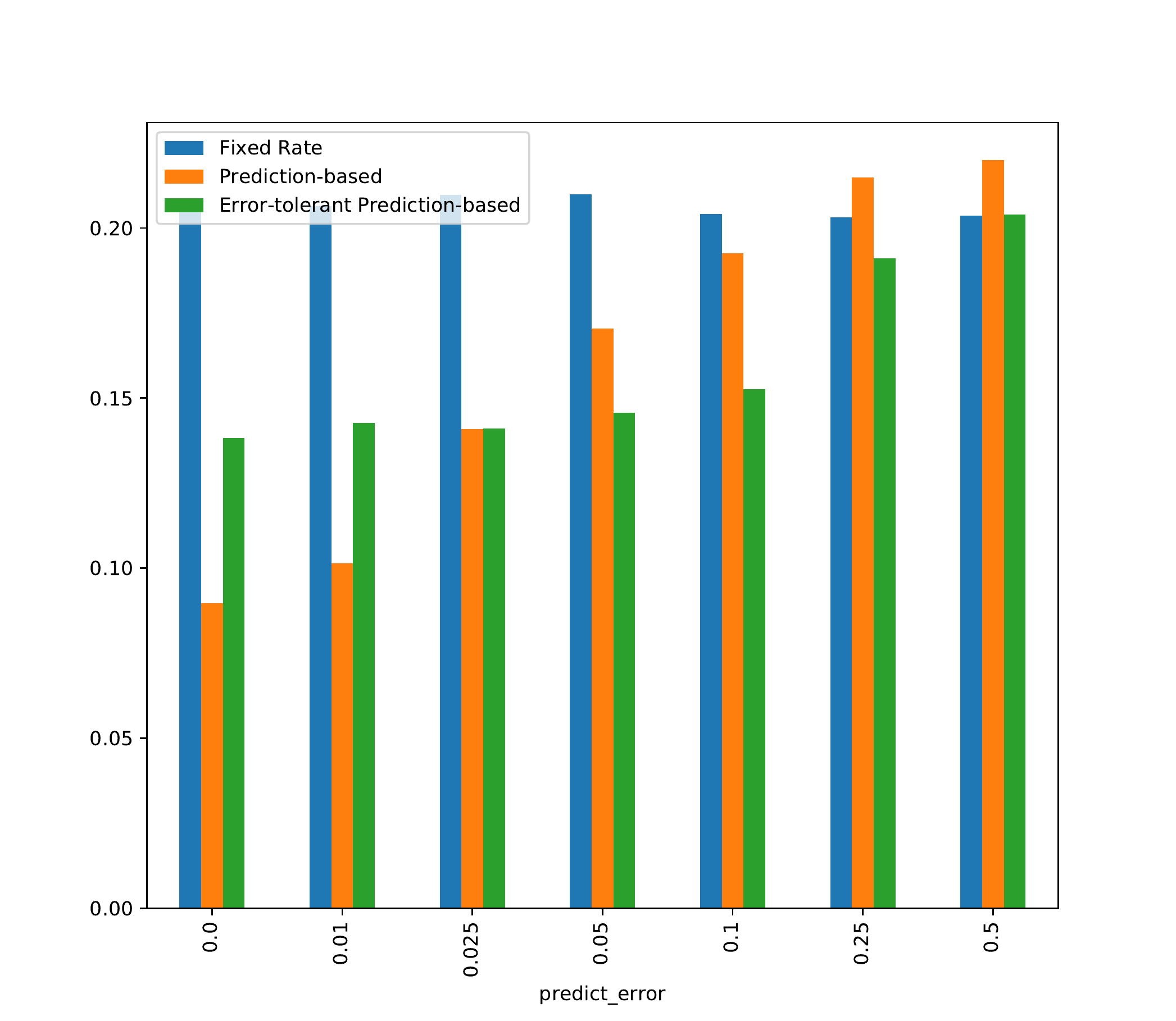}
\caption{\label{fig:prediction}Performance impact (slowdown over a failure-free
execution) of different prediction models. Y-axis: $T_{exe} - T_{ideal} ) /
T_{ideal}$, X-axis: MTTF prediction error.}
\end{figure}

Figure~\ref{fig:prediction} shows the results of a simulated experiment, carried out
under the following assumptions: the failure rate is set so that the baseline
(not reported) has an average slowdown of 1; checkpoint time is around 1.5-2\%
of the application (ideal) execution time; restart overhead is about 15-20\% of
the application (ideal) execution time. Figure~\ref{fig:prediction} reports,
the average overhead due to reliability (i.e., $ O = T_{exe} - T_{ideal} ) /
T_{ideal} $),  as a function of the MTTF prediction error.
For the execution of the experiments, we generated random workloads over a long
time window (around 200$\times$ the execution time of a single application).
Experiments have been repeated 20 times, taking median values.
The fixed-rate policy was configured with a checkpointing interval set at 20$\times$
the checkpointing time.  As shown in the figure,
prediction-based policies are more effective when an effective prediction
(relative prediction error is shown on the $y$ axis) is available.
The error-tolerant policy instead, can come into play in case the reliability
models cannot guarantee prediction errors lower than $2.5\%$.

\section{Weather forecasting use-case}
\label{usecases-sec}
RECIPE identified a number of real-world high performance applications demonstrating the emerging class of requirements
addressed by the project in terms of heterogeneity, fault-tolerance, and time constraints.
In particular, the RECIPE use cases include applications in the areas of
weather forecasting, subsoil properties identification, and bio-medical big-data applications.
The ultimate objective in RECIPE is to perform a demonstration on both industry-grade, pre-Exascale systems
and on emerging deeply heterogeneous technologies, in order to measure the degree of success of the project.

\begin{figure}[t!]
	\includegraphics[width=\columnwidth]{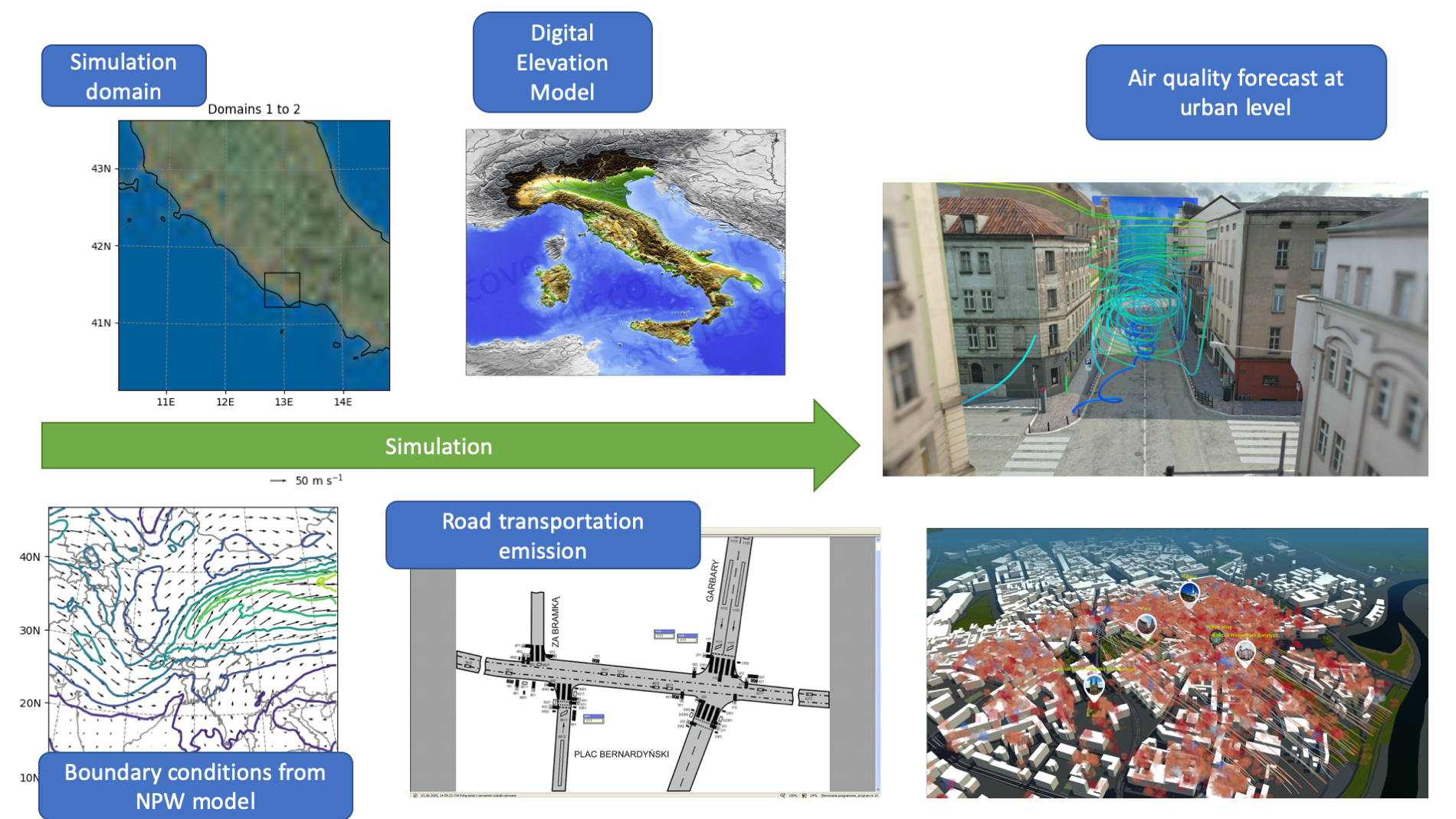}
	\centering
	\caption[Highlevel overview of UrbanAir application]{High level representation 
	of the UrbanAir application.\newline Different input parameters, e.g. digital
	elevation model, road transporation emission, mesoscale weather prediction
	are used to model air quality over urban areas, which are then visualized at
	city and street scale.}
	\label{fig:uc2_psnc}
\end{figure}

In the weather forecasting area, two application are addressed: the first one, UrbanAir,
is a multiscale model that allows for weather prediction, wind farm
optimizations and prediction of air quality over complex urban
areas~\cite{eulag_uq}, see Figure~\ref{fig:uc2_psnc} for example. 
The second one is a flood prediction model, supported by on-field devices.
\par
The UrbanAir uses two models - WRF for weather prediction at mesoscale to local
level, and EULAG for detailed wind prediction and air quality modelling at
local to street level. Both models are HPC application written in FORTRAN that 
uses MPI parallel programming paradigm. Both models have been coupled to each
other so that prediction results of the former are used as input parameters to
the latter. As for the EULAG model, there are some very small parts of code, 
called kernels, available for the GPU accelerators, though they are not part of
the production code yet.
\par
Within the RECIPE context, the UrbanAir application will benefit from fault
tolerance and time constraints features in particular. Weather prediction for
emergency scenarios, such as flood prediction, requires timely analysis. This
will be achieved by combining different methods of speeding up simulations
(e.g. coarser domain, longer time-step, different physics and parametrization
schemes) and RECIPE tools to obey required time regime.
For the long-running scenarios, e.g. weather prediction for wind farm
optimization, the key requirements is to provide reliable forecast under given
time constraint and hardware/software failures. The UrbanAir application
already supports restarting simulation from a given point in time, which will
be coupled to fault tolerance solution provided within the project.
\par
In the emergency scenarios, such as flood prediction, 15
minutes regime has been defined within which NWP prediction has to be delivered. 
To support DSOs operations, NWP prediction has
to be provided within 24 hours. The application performance may be sensitive to
memory layout or task placement across HPC machine. In such situation, the MBPTA 
helps to analyze the worst case execution, so that proper parameters values of 
the simulation may be selected, as well as the highest possible resolution 
of the simulation domain. In this way, NWP prediction is delivered with the
highest quality of results possible under given time constraints.

\section{Conclusions}
\label{conclusions-sec}

In this paper, we presented an overview of the technological challenges emerging from the coupling of deeply heterogeneous HPC architectures in Exascale systems with emerging application domains characterised by QoS requirements.
We identified proactive and reactive reliability mechanisms as necessary to support this scenario, and introduced the key innovations brought by the RECIPE H2020 FETHPC project to address is.
Such technologies span
run-time management, heterogeneous computing architectures,
HPC memory/interconnection infrastructures, thermal modelling, reliability, programming models, and timing analysis.

We provided an initial assessment of the impact of reliability techniques such as checkpoint/restore mechanisms when coupled with fault prediction models.
During the second half of the RECIPE project, we will work towards the integration of such mechanisms in the context of the disaggregated management of the heterogeneous computing resources, to provide a flexible environment to run applications with widely different requirements.

\section*{Acknowledgement}
The activities described in this article received funding from the European Union's Horizon 2020 research
and innovation programme under the FETHPC grant agreement no. 801137
RECIPE: REliable power and time-ConstraInts-aware Predictive management of heterogeneous Exascale systems.

% \bibliographystyle{plain}
% \bibliography{biblio}

% references section

% can use a bibliography generated by BibTeX as a .bbl file
% BibTeX documentation can be easily obtained at:
% http://www.ctan.org/tex-archive/biblio/bibtex/contrib/doc/
% The IEEEtran BibTeX style support page is at:
% http://www.michaelshell.org/tex/ieeetran/bibtex/
\bibliographystyle{IEEEtran}
% argument is your BibTeX string definitions and bibliography database(s)
\bibliography{IEEEabrv,biblio}

% Generated by IEEEtran.bst, version: 1.14 (2015/08/26)
\begin{thebibliography}{10}
\providecommand{\url}[1]{#1}
\csname url@samestyle\endcsname
\providecommand{\newblock}{\relax}
\providecommand{\bibinfo}[2]{#2}
\providecommand{\BIBentrySTDinterwordspacing}{\spaceskip=0pt\relax}
\providecommand{\BIBentryALTinterwordstretchfactor}{4}
\providecommand{\BIBentryALTinterwordspacing}{\spaceskip=\fontdimen2\font plus
\BIBentryALTinterwordstretchfactor\fontdimen3\font minus
  \fontdimen4\font\relax}
\providecommand{\BIBforeignlanguage}[2]{{%
\expandafter\ifx\csname l@#1\endcsname\relax
\typeout{** WARNING: IEEEtran.bst: No hyphenation pattern has been}%
\typeout{** loaded for the language `#1'. Using the pattern for}%
\typeout{** the default language instead.}%
\else
\language=\csname l@#1\endcsname
\fi
#2}}
\providecommand{\BIBdecl}{\relax}
\BIBdecl

\bibitem{sra2017}
J.-P. Panziera \emph{et~al.}, ``{Strategic Research Agenda 2017},'' ETP4HPC,
  Tech. Rep., Nov 2017.

\bibitem{fortissimo}
B.~Koller \emph{et~al.}, ``Towards an environment to deliver high performance
  computing to small and medium enterprises,'' in \emph{Sustained Simulation
  Performance 2015}.\hskip 1em plus 0.5em minus 0.4em\relax Cham: Springer
  International Publishing, 2015, pp. 41--50.

\bibitem{agosta2018managing}
\BIBentryALTinterwordspacing
G.~Agosta, W.~Fornaciari, G.~Massari, A.~Pupykina, F.~Reghenzani, and
  M.~Zanella, ``{Managing Heterogeneous Resources in {HPC} Systems},'' in
  \emph{Proc. of PARMA-DITAM '18}.\hskip 1em plus 0.5em minus 0.4em\relax ACM,
  2018, pp. 7--12. [Online]. Available:
  \url{http://doi.acm.org/10.1145/3183767.3183769}
\BIBentrySTDinterwordspacing

\bibitem{flich2016mango}
J.~Flich \emph{et~al.}, ``{Enabling {HPC} for {QoS}-sensitive applications: The
  {MANGO} approach},'' in \emph{2016 Design, Automation Test in Europe
  Conference Exhibition (DATE)}, March 2016, pp. 702--707.

\bibitem{flich2018mango}
\BIBentryALTinterwordspacing
------, ``Exploring manycore architectures for next-generation {HPC} systems
  through the {MANGO} approach,'' \emph{Microprocessors and Microsystems},
  vol.~61, pp. 154 -- 170, 2018. [Online]. Available:
  \url{http://www.sciencedirect.com/science/article/pii/ S0141933118300243}
\BIBentrySTDinterwordspacing

\bibitem{samos2019}
G.~Massari, A.~Pupykina, G.~Agosta, and W.~Fornaciari, ``Predictive resource
  management for next-generation high-performance computing heterogeneous
  platforms,'' in \emph{Proceedings of the 18th International Conference on
  Embedded Computer Systems: Architectures, Modeling, and Simulation
  (SAMOS'19)}, Jul 2019.

\bibitem{pupykina2017optimizing}
A.~Pupykina and G.~Agosta, ``{Optimizing Memory Management in Deeply
  Heterogeneous {HPC} Accelerators},'' in \emph{2017 46th Int'l Conf on
  Parallel Processing Workshops (ICPPW)}, Aug 2017, pp. 291--300.

\bibitem{flich2015mango}
J.~Flich \emph{et~al.}, ``{The {MANGO} {FET-HPC} Project: An overview},'' in
  \emph{IEEE 18th Int'l Conf on Computational Science and Engineering
  (CSE)}.\hskip 1em plus 0.5em minus 0.4em\relax IEEE, 2015, pp. 351--354.

\bibitem{Zanella2018back}
\BIBentryALTinterwordspacing
M.~Zanella, G.~Massari, A.~Galimberti, and W.~Fornaciari, ``Back to the future:
  Resource management in post-cloud solutions,'' in \emph{Proceedings of the
  Workshop on INTelligent Embedded Systems Architectures and Applications},
  ser. INTESA '18.\hskip 1em plus 0.5em minus 0.4em\relax New York, NY, USA:
  ACM, 2018, pp. 33--38. [Online]. Available:
  \url{http://doi.acm.org/10.1145/3285017.3285028}
\BIBentrySTDinterwordspacing

\bibitem{Fornaciari2018reliable}
\BIBentryALTinterwordspacing
W.~Fornaciari, G.~Agosta, D.~Atienza, C.~Brandolese, L.~Cammoun, L.~Cremona,
  A.~Cilardo, A.~Farres, J.~Flich, C.~Hernandez, M.~Kulchewski, S.~Libutti,
  J.~M. Mart\'{\i}nez, G.~Massari, A.~Oleksiak, A.~Pupykina, F.~Reghenzani,
  R.~Tornero, M.~Zanella, M.~Zapater, and D.~Zoni, ``Reliable power and
  time-constraints-aware predictive management of heterogeneous exascale
  systems,'' in \emph{Proceedings of the 18th International Conference on
  Embedded Computer Systems: Architectures, Modeling, and Simulation}, ser.
  SAMOS '18.\hskip 1em plus 0.5em minus 0.4em\relax New York, NY, USA: ACM,
  2018, pp. 187--194. [Online]. Available:
  \url{http://doi.acm.org/10.1145/3229631.3239368}
\BIBentrySTDinterwordspacing

\bibitem{Farm2011}
C.~L. Chou and R.~Marculescu, ``{FARM}: Fault-aware resource management in
  {NoC}-based multiprocessor platforms,'' in \emph{2011 Design, Automation Test
  in Europe}, March 2011, pp. 1--6.

\bibitem{Haghbayan2016}
M.~H. Haghbayan, A.~Miele, A.~M. Rahmani, P.~Liljeberg, and H.~Tenhunen, ``A
  lifetime-aware runtime mapping approach for many-core systems in the dark
  silicon era,'' in \emph{2016 Design, Automation Test in Europe Conference
  (DATE)}, March 2016, pp. 854--857.

\bibitem{Huang2010}
L.~Huang and Q.~Xu, ``Characterizing the lifetime reliability of manycore
  processors with core-level redundancy,'' in \emph{2010 IEEE/ACM International
  Conference on Computer-Aided Design (ICCAD)}, Nov 2010, pp. 680--685.

\bibitem{Warm2016}
P.~Mercati, F.~Paterna, A.~Bartolini, L.~Benini, and T.~Rosing, ``{WARM}:
  Workload-aware reliability management in {L}inux/{A}ndroid,'' \emph{IEEE
  Trans on CAD of Integrated Circuits and Systems}, 2016.

\bibitem{bib:euroexa}
``https://euroexa.eu.''

\bibitem{bib:intelstratix}
``https://www.altera.com/products/sip/memory/stratix-10-mx/overview.html.''

\bibitem{bib:mango}
``http://www.mango-project.eu.''

\bibitem{cilardo15}
A.~Cilardo and L.~Gallo, ``Interplay of loop unrolling and multidimensional
  memory partitioning in {HLS},'' in \emph{Proceedings of the 2015 Design,
  Automation and Test in Europe Conference and Exhibition}, ser. DATE
  '15.\hskip 1em plus 0.5em minus 0.4em\relax San Jose, CA, USA: EDA
  Consortium, 2015, pp. 163--168.

\bibitem{paranjape2012heterogeneous}
K.~Paranjape, S.~Hebert, and B.~Masson, ``Heterogeneous computing in the cloud:
  Crunching big data and democratizing {HPC} access for the life sciences,''
  Intel Corporation, Tech. Rep., 2010.

\bibitem{sarkar10}
S.~Sarkar, T.~Majumder, A.~Kalyanaraman, and P.~Pande, ``Hardware accelerators
  for biocomputing: A survey,'' in \emph{Circuits and Systems (ISCAS),
  Proceedings of 2010 IEEE International Symposium on}, May 2010, pp.
  3789--3792.

\bibitem{eth2017}
``{IEEE} standard for {E}thernet - {A}mendment 10,'' \emph{IEEE Std
  802.3bs-2017}, pp. 1--372, Dec 2017.

\bibitem{ib2001}
G.~F. Pfister, ``Aspects of the {I}nfini{B}and architecture,'' in
  \emph{Proceedings 42nd IEEE Symposium on Foundations of Computer Science},
  Oct 2001, pp. 369--371.

\bibitem{op2015}
M.~S. Birrittella, M.~Debbage, R.~Huggahalli, J.~Kunz, T.~Lovett, T.~Rimmer,
  K.~D. Underwood, and R.~C. Zak, ``Intel® {O}mni-{P}ath architecture:
  Enabling scalable, high performance fabrics,'' in \emph{2015 IEEE 23rd Annual
  Symposium on High-Performance Interconnects}, Aug 2015, pp. 1--9.

\bibitem{bib:ibta}
``https://www.infinibandta.org/infiniband-roadmap/.''

\bibitem{zoni2019alldigital}
D.~{Zoni}, L.~{Cremona}, and W.~{Fornaciari}, ``All-digital energy-constrained
  controller for general-purpose accelerators and {CPUs},'' \emph{IEEE Embedded
  Systems Letters}, 2019, doi:10.1109/LES.2019.2914136.

\bibitem{Reghenzani2019pWCET}
\BIBentryALTinterwordspacing
F.~Reghenzani, G.~Massari, W.~Fornaciari, and A.~Galimberti,
  ``Probabilistic-{WCET} reliability: On the experimental validation of {EVT}
  hypotheses,'' in \emph{Proceedings of the International Conference on
  Omni-Layer Intelligent Systems}, ser. COINS '19.\hskip 1em plus 0.5em minus
  0.4em\relax New York, NY, USA: ACM, 2019, pp. 229--234. [Online]. Available:
  \url{http://doi.acm.org/10.1145/3312614.3312660}
\BIBentrySTDinterwordspacing

\bibitem{reghenzani2018chronovise}
F.~Reghenzani, G.~Massari, and W.~Fornaciari, ``chronovise: {M}easurement-based
  probabilistic timing analysis framework,'' \emph{J. Open Source Software},
  vol.~3, no.~28, p. 711, 2018.

\bibitem{8585132}
F.~{Reghenzani}, G.~{Massari}, and W.~{Fornaciari}, ``The misconception of
  exponential tail upper-bounding in probabilistic real-time,'' \emph{IEEE
  Embedded Systems Letters}, 2018, doi:10.1109/LES.2018.2889114.

\bibitem{EVTECRTS}
L.~{Cucu-Grosjean et al.}, ``Measurement-based probabilistic timing analysis
  for multi-path programs,'' in \emph{ECRTS}, 2012.

\bibitem{MBPTACV}
\BIBentryALTinterwordspacing
J.~Abella, M.~Padilla, J.~D. Castillo, and F.~J. Cazorla, ``Measurement-based
  worst-case execution time estimation using the coefficient of variation,''
  \emph{ACM Trans. Des. Autom. Electron. Syst.}, vol.~22, no.~4, Jun. 2017.
  [Online]. Available: \url{https://doi.org/10.1145/3065924}
\BIBentrySTDinterwordspacing

\bibitem{bib:trinity}
``https://lanl.gov/projects/trinity/specifications.php.''

\bibitem{bib:marenostrum}
``https://www.bsc.es/marenostrum/marenostrum/technical-information.''

\bibitem{bib:titan}
``https://www.olcf.ornl.gov/olcf-resources/compute-systems/titan/.''

\bibitem{Bellasi2015Effective}
\BIBentryALTinterwordspacing
P.~Bellasi, G.~Massari, and W.~Fornaciari, ``Effective runtime resource
  management using linux control groups with the {BarbequeRTRM} framework,''
  \emph{ACM Trans. Embed. Comput. Syst.}, vol.~14, no.~2, pp. 39:1--39:17, Mar.
  2015. [Online]. Available: \url{http://doi.acm.org/10.1145/2658990}
\BIBentrySTDinterwordspacing

\bibitem{3d-ice}
A.~{Sridhar}, A.~{Vincenzi}, M.~{Ruggiero}, T.~{Brunschwiler}, and
  D.~{Atienza}, ``{3D-ICE}: Fast compact transient thermal modeling for {3D
  ICs} with inter-tier liquid cooling,'' in \emph{2010 IEEE/ACM International
  Conference on Computer-Aided Design (ICCAD)}, Nov 2010, pp. 463--470.

\bibitem{gem5X}
Y.~M. Qureshi, W.~A. Simon, M.~Zapater~Sancho, K.~Olcoz, and D.~Atienza~Alonso,
  ``{G}em5-{X}: A {G}em5-based system level simulation framework to optimize
  many-core platforms,'' \emph{Proceedings of the 27th High Performance
  Computing Symposium (HPC 2019)}, p.~12, 2019.

\bibitem{faultprediction}
A.~Gainaru, F.~Cappello, M.~Snir, and W.~Kramer, ``Fault prediction under the
  microscope: A closer look into {HPC} systems,'' in \emph{SC '12: Proc. of the
  Int. Conference on High Performance Computing, Networking, Storage and
  Analysis}, Nov 2012, pp. 1--11.

\bibitem{Egwutuoha2013}
I.~Egwutuoha, D.~Levy, B.~Selic, and S.~Chen, ``A survey of fault tolerance
  mechanisms and checkpoint/restart implementations for high performance
  computing systems,'' \emph{The Journal of Supercomputing}, vol.~65, 09 2013.

\bibitem{Lee2017}
K.~Lee and S.~S. Wong, ``{Fault-Tolerant FPGA with Column-Based Redundancy and
  Power Gating Using RRAM},'' \emph{IEEE Transactions on Computers}, vol.~66,
  no.~6, pp. 946--956, 2017.

\bibitem{Cheatham2006}
J.~A. Cheatham, J.~M. Emmert, and S.~Baumgart, ``{A survey of fault tolerant
  methodologies for FPGAs},'' \emph{ACM Transactions on Design Automation of
  Electronic Systems}, vol.~11, no.~2, pp. 501--533, 2006.

\bibitem{Parris2011}
M.~G. Parris, C.~A. Sharma, and R.~F. Demara, ``{Progress in autonomous fault
  recovery of Field Programmable Gate Arrays},'' \emph{ACM Computing Surveys},
  vol.~43, no.~4, 2011.

\bibitem{heatsinkplug}
A.~Iranfar, F.~Terraneo, W.~A. Simon, L.~Dragic, I.~Pilji, M.~Zapater~Sancho,
  W.~Fornaciari, M.~Kovac, and D.~Atienza~Alonso, ``Thermal characterization of
  next-generation workloads on heterogeneous {MPSoCs},'' 2017.

\bibitem{zoni2015modeling}
\BIBentryALTinterwordspacing
D.~Zoni and W.~Fornaciari, ``Modeling {DVFS} and power-gating actuators for
  cycle-accurate {NoC}-based simulators,'' \emph{J. Emerg. Technol. Comput.
  Syst.}, vol.~12, no.~3, pp. 27:1--27:24, Sep. 2015. [Online]. Available:
  \url{http://doi.acm.org/10.1145/2751561}
\BIBentrySTDinterwordspacing

\bibitem{IranfarDATE2020}
A.~Iranfar, F.~Terraneo, G.~Csordas, M.~Zapater, W.~Fornaciari, and D.~Atienza,
  ``Dynamic thermal management with proactive fan speed control through
  reinforcement learning,'' \emph{[Proceedings Design, Automation and Test in
  Europe Conference and Exhibition (DATE)]}, p.~6, 2020.

\bibitem{terraneo2019open}
F.~Terraneo, A.~Leva, and W.~Fornaciari, ``An open-hardware platform for mpsoc
  thermal modeling,'' in \emph{International Conference on Embedded Computer
  Systems}.\hskip 1em plus 0.5em minus 0.4em\relax Springer, 2019.

\bibitem{Stabilizer}
\BIBentryALTinterwordspacing
C.~Curtsinger and E.~D. Berger, ``{STABILIZER}: Statistically sound performance
  evaluation,'' \emph{SIGARCH Comput. Archit. News}, vol.~41, no.~1, pp.
  219--228, Mar. 2013. [Online]. Available:
  \url{http://doi.acm.org/10.1145/2490301.2451141}
\BIBentrySTDinterwordspacing

\bibitem{SWRAND}
L.~Kosmidis, C.~Curtsinger, E.~Quiones, J.~Abella, E.~Berger, and F.~J.
  Cazorla, ``Probabilistic timing analysis on conventional cache designs,'' in
  \emph{2013 Design, Automation Test in Europe Conference Exhibition (DATE)},
  March 2013, pp. 603--606.

\bibitem{fwi1}
J.~Kormann, J.~Rodríguez, N.~Gutiérrez, M.~Ferrer, O.~Rojas, J.~De~la Puente,
  M.~Hanzich, and J.~María~Cela, ``Toward an automatic full-wave inversion:
  Synthetic study cases,'' \emph{The Leading Edge}, vol.~35, pp. 1047--1052, 12
  2016.

\bibitem{fwi2}
M.~Hanzich, J.~Kormann, N.~Gutiérrez, J.~Rodríguez, J.~De~la Puente, and
  J.~María~Cela, ``Developing full waveform inversion using hpc frameworks:
  Bsit,'' in \emph{EAGE Workshop on High Performance Computing for Upstream},
  09 2014.

\bibitem{Fusi2020}
\BIBentryALTinterwordspacing
M.~Fusi, F.~Mazzocchetti, A.~Farres, L.~Kosmidis, R.~Canal, F.~J. Cazorla, and
  J.~Abella, ``On the use of probabilistic worst-case execution time estimation
  for parallel applications in high performance systems,'' \emph{Mathematics},
  vol.~8, no.~3, pp. 314--335, Mar 2020. [Online]. Available:
  \url{http://dx.doi.org/10.3390/math8030314}
\BIBentrySTDinterwordspacing

\bibitem{eulag_uq}
D.~W. Wright, R.~A. Richardson, W.~Edeling, J.~Lakhlili, R.~C. Sinclair,
  V.~Jacauskas, D.~Suleimenova, B.~Bosak, M.~Kulczewski, T.~Piontek, P.~Kopta,
  I.~Chirca, H.~Arabnejad, O.~O. Luk, O.~Hoenen, J.~Weglarz, D.~Crommelin, and
  D.~Groen, ``Building confidence in simulation: Application of easyvvuq,''
  submitted to Journal of Advanced Theory and Simulations on 12/12/2019.

\end{thebibliography}
%
% <OR> manually copy in the resultant .bbl file
% set second argument of \begin to the number of references
% (used to reserve space for the reference number labels box)
% \begin{thebibliography}{1}
%
% \bibitem{IEEEhowto:kopka}
% H.~Kopka and P.~W. Daly, \emph{A Guide to \LaTeX}, 3rd~ed.\hskip 1em plus
%   0.5em minus 0.4em\relax Harlow, England: Addison-Wesley, 1999.
%
% \end{thebibliography}

% that's all folks
\end{document}